\documentclass[twocolumn, pra,
showpacs,superscriptaddress]{revtex4}
\usepackage{graphicx}
\usepackage{dcolumn}
\usepackage{bm}
\usepackage{amsmath}

\begin{document}

\title{Ground-state properties of hard-core anyons in one-dimensional optical lattices}
\author{Yajiang Hao}
\affiliation{Department of Physics, University of Science and
Technology Beijing, Beijing 100083, China}
\affiliation {Beijing
National Laboratory for Condensed Matter Physics, Institute of
Physics, Chinese Academy of Sciences, Beijing 100080, China}

\author{Yunbo Zhang}
\affiliation{Department of Physics and Institute of Theoretical
Physics, Shanxi University, Taiyuan 030006, China}
\author{Shu Chen}
\email{schen@aphy.iphy.ac.cn} \affiliation{Beijing National
Laboratory for Condensed Matter Physics, Institute of Physics,
Chinese Academy of Sciences, Beijing 100080, China}
\date{\today}

\begin{abstract}
We investigate the ground-state properties of anyons confined in
one-dimensional optical lattices with a weak harmonic trap using the
exact numerical method based on Jordan-Wigner transformation. It is
shown that in the Bose limit ($\chi =1$) and Fermi limit ($\chi=0$)
the momentum distributions are symmetric but in between they are
asymmetric. It turns out that the origin of asymmetry comes from the
fractional statistics that anyons obey. The occupation distribution
and the modulus of natural orbitals show crossover behaviors from
the Bose limit to the Fermi limit.
\end{abstract}
 \maketitle
% 05.30.Pr Fractional statistics systems
% 03.75.Hh Static properties of condensates; thermodynamical, statistical and structural properties.
% 05.30.Fk Fermion systems and electron gas
% 05.30.Jp Boson systems
% 67.40.Db Quantum statistical theory; groundstate, elementary excitations

%\preprint{APS/123-QED}

\narrowtext

\section{introduction}

The physical systems with fractional statistics have been a subject
of great interest in past decades and been intensively studied for
two-dimensional systems \cite{anyons,Laughlin,Halperin,Camino}. For
instance, the elementary excitations of a fractional quantum Hall
(FQH) liquid satisfy fractional statistics. This has been observed
in two-dimensional electron gas and anyon has become an important
concept in studying the FQH effect \cite{Wilczek,anyons,Haldane}. As
a natural generalization of the Bose and Fermi gas, anyon gas has
also found application in various one-dimensional (1D) systems
\cite{Haldane,WangZD,Kundu99,Girardeau06}.  Despite no explicitly
experimental proof of the realization of 1D anyon gas, many
theorists have dedicated to study them
\cite{Kundu99,Girardeau06,Batchelor06PRL,Patu07,Batchelor07,
Batchelor06PRB,Calabrese,anyonTG,
Patu08,Cabra,Campo,HaoPRA78,Amico,Batchelor}. Currently anyons also
stimulated intensive research on topological quantum computation
because the statistical properties are closely related to the
topological order. Particularly, by controlling exchange interaction
between pairs of neutral atoms in optical lattice, Anderlini
\emph{et al.} \cite{SWAP} realized the key operation in quantum
information processing, i.e., the quantum SWAP gate (which control
the states interchange between two qubits). Cold atoms in optical
lattice are also proposed to create, manipulate, and test anyons
\cite{PZoller,Aguado,Jiang}.

By tightly confining the particle motion in two directions to zero
point oscillations 1D quantum gas is obtained
\cite{Ketterler,Paredes,Toshiya}, where the radial degrees of
freedom are frozen and the quantum gas is effectively described by a
1D model \cite{Olshanii}. Experimentally, by means of anisotropic
magnetic trap or two-dimensional optical lattice potentials, a 1D
Bose gas in the strongly correlated Tonks-Girardeau (TG) regime can
be achieved \cite{Paredes,Toshiya}. The TG gas has been shown to
display the "fermionized" character in many aspects. For example, it
has the same density distribution and thermodynamic behavior as the
free fermion \cite{Girardeau,Hao06,Hao07,Zoellner,Deuretzbacher}.
However, the off-diagonal density matrices and the momentum
distributions exhibit quite different behaviors due to the different
permutation symmetries of the Bose and Fermi wave functions
\cite{Lenard,Vaidya}. The momentum distribution of fermions shows
typical oscillations but that of bosons shows the structure of
single peak. Both of them are symmetric about the zero momentum. For
the anyon gas satisfying fractional statistics, however the momentum
distribution is shown to be asymmetric when the statistical
parameters deviate from the Bose and Fermi limit
\cite{HaoPRA78,anyonTG,Patu08,Campo}, and with the change in
statistical parameter the distribution evolves from a Bose
distribution to a Fermi one and vice versa.

While most of the theoretical works on the 1D anyon gas focus on the
continuum system,  in this paper we investigate the ground state of
a 1D anyonic system confined in optical lattice with weak harmonic
trap in the hard core limit. Although the hard-core bosons have been
studied extensively and intensively for both homogeneous continuum
system and lattice system, no result was given for the hard-core
anyons (HCAs) in optical lattice. We shall study how the fractional
statistics affect the ground-state properties, such as the momentum
distribution. By extending the exact numerical method originally
used to treat hard-core bosons by Rigol and Muramatsu \cite{MRigol}
to deal with the hard-core anyons, we evaluate the exact
one-particle Green's function of the ground state, with which the
reduced one body density matrix (ROBDM) and thus the momentum
distribution can be obtained exactly for different statistical
parameters. This method is based on Jordan-Wigner transformation and
has been applied in the investigation of hard core bosons confined
in optical lattice for both ground state and dynamics. It turns out
to be very efficient to study the universal behaviors of the system
with arbitrary confining potentials combined with one-dimensional
optical lattices. The properties of anyonic statistics shall be
displayed in the momentum distribution and with the change in
statistics parameter the system exhibits the Bose statistics, Fermi
statistics, and the fractional statistics in between. The
mathematical origin of the asymmetry can be traced back to the
reduced one body density matrix.

The paper is organized as follows. In Sec. II, we give a brief
review of 1D anyonic model and introduce the numerical method. In
Sec. III, we present the momentum distributions, ROBDM, and
occupation for different statistics parameter and filling numbers. A
brief summary is given in Sec. IV.

\section{formulation of the model and method}

The second quantized Hamiltonian of the hard-core anyons confined in
optical lattice of\ $L$ sites with a weak harmonic trap takes the
form of
\begin{equation}
H_{\text{HCA}}=-t\sum_{l=1}^L\left( a_{l+1}^{\dagger
}a_l+\text{H.C.}\right) +\sum_{l=1}^LV_ln_l  \label{Ham}
\end{equation}
with the lattice dependent external potential
\[
V_l=V_0(l-(L+1)/2)^2,
\]
where $V_0$ denotes the strength of the harmonic trap. The anyonic
creation operator $a_l^{\dagger }$ and annihilation operator $a_j$
satisfy the generalized commutation relations
\begin{eqnarray}
a_ja_l^{\dagger } &=&\delta _{jl}-e^{-i\chi \pi \epsilon
(j-l)}a_l^{\dagger }a_j, \nonumber\\
a_ja_l &=&-e^{i\chi \pi \epsilon (j-l)}a_la_j
\end{eqnarray}
for $j\neq l$ with the addition of hard-core condition $a_l^2=a_l^{\dagger
2}=0$ and $\left\{ a_l,a_l^{\dagger }\right\} =1$. The sign function $%
\epsilon (x)$ gives -1, 0, or 1 depending on whether $x$ is
negative, zero, or positive. The parameter $\chi $ is related with
fractional statistics and will be restricted in the regime of
$\left[ 0,1 \right] $ in the present paper. Particularly, $\chi =0$
and $1$ correspond to Fermi statistics and Bose statistics,
respectively. In the Hamiltonian the particle number operator
$n_l=a_l^{\dagger }a_l$, and $t$ denotes the hopping between the
nearest neighbor sites.

This model can be solved exactly by the generalized Jordan-Wigner
transformation,
\begin{eqnarray}
a_j &=&\exp \left( i\chi \pi \sum_{1\leq s<j}f_s^{\dagger }f_s\right) f_j,\nonumber \\
a_j^{\dagger } &=&f_j^{\dagger }\exp \left( -i\chi \pi \sum_{1\leq
s<j}f_s^{\dagger }f_s\right) ,
\end{eqnarray}
where $f_j^{\dagger }$ and $f_j$ are creation and annihilation
operators for spinless fermions. The hard-core anyonic Hamiltonian
with $N$ anyons can be mapped onto the noninteracting fermionic
system for $N_F$ fermions ($N_F=N$),
\begin{eqnarray}
H_F=-t\sum_{l=1}^L\left( f_{l+1}^{\dagger }f_l+\text{H.C.}\right)
+\sum_{l=1}^LV_ln_l
\end{eqnarray}
with fermionic particle number operator $n_l=f_l^{\dagger }f_l$.
While the eigen problem for Hamiltonian $H_F$ for vanishing harmonic
potential $V_l=0$ can be obtained easily through Fourier
transformation, we investigate here the situation with weak harmonic
trap and Rigol-Muramatsu method \cite{MRigol} should be a good
choice. Using the above transformation the one-particle Green's
function of hard-core anyon is formulated as
\begin{eqnarray}
G_{jl} &=&\left\langle \Psi _{\text{HCA}}^G\left| a_ja_l^{\dagger
}\right| \Psi
_{\text{HCA}}^G\right\rangle   \nonumber \\
&=&\left\langle \Psi _F^G\left| e^{  i\chi \pi \sum_\beta
^{j-1}f_\beta ^{\dagger }f_\beta } f_jf_l^{\dagger }e^{ -i\chi \pi
\sum_\gamma ^{l-1}f_\gamma ^{\dagger }c_\gamma  }\right| \Psi
_F^G\right\rangle   \nonumber \\
&=&\left\langle \Psi _F^A|\Psi _F^B\right\rangle \label{GF}
\end{eqnarray}
with
\begin{eqnarray*}
\left\langle \Psi _F^A\right|  &=&\left[ f_j^{\dagger }\exp \left(
-i\chi \pi \sum_\beta ^{j-1}f_\beta ^{\dagger }f_\beta \right)
\left| \Psi
_F^G\right\rangle \right] ^{\dagger }, \\
\left| \Psi _F^B\right\rangle  &=&f_l^{\dagger }\exp \left( -i\chi
\pi \sum_\gamma ^{l-1}f_\gamma ^{\dagger }f_\gamma \right) \left|
\Psi _F^G\right\rangle .
\end{eqnarray*}
$\left| \Psi _{\text{HCA}}^G\right\rangle $ is the ground state of
hard core anyonic system and $\left| \Psi _F^G\right\rangle $ is the
ground state of free spinless fermionic system. The Green's function
can be obtained by constructing the many-particle ground state of
fermions with the eigenstates of single-particle,
\[
\left| \alpha \right\rangle =c_\alpha ^{\dagger }\left| 0\right\rangle
=\sum_l\varphi _\alpha \left( l\right) f_l^{\dagger }\left| 0\right\rangle ,
\]
where $\alpha $ means the $\alpha $th state and $l$ means the $l$th
site. The many body ground state of $N_f$ free spinless Fermions
takes the following form:
\begin{eqnarray}
\left| \Psi _F^G\right\rangle =c_1^{\dagger }c_2^{\dagger }\cdots
c_{N_f}^{\dagger }\left| 0\right\rangle
=\prod_{n=1}^{N_f}\sum_{l=1}^LP_{ln}f_l^{\dagger }\left|
0\right\rangle
\end{eqnarray}
with $P_{ln}=\varphi _n\left( l\right) $, which can be expressed as
an $L\times N_f$ matrix {\bf P}. After an easy evaluation the state
$\left| \Psi _F^A\right\rangle $ reads
\[
\left| \Psi _F^A\right\rangle =\prod_{n=1}^{N_f+1}\sum_{l=1}^LP_{ln}^{\prime
A}f_l^{\dagger }\left| 0\right\rangle
\]
with
\begin{eqnarray*}
P_{ln}^{\prime A} &=&\exp \left( -i\chi \pi \right) P_{ln}
\begin{array}{lll}
&  &
\end{array}
\text{for }l\leq j-1 \\
P_{ln}^{\prime A} &=&P_{ln}
\begin{array}{llllllllllll}
&  &  &  &  &  &  &  &  &  &  &
\end{array}
\text{for }l\geq j
\end{eqnarray*}
for $n\leq N_f$, and $P_{jN_f+1}^{\prime A}=1$ and $P_{lN_f+1}^{\prime A}=0$ $%
\left( l\neq j\right) $. The state $\left| \Psi _F^B\right\rangle $
has the same form with the replacement of $j$ by $l$. The Green's
function is a
determinant dependent on the $L\times \left( N_f+1\right) $ matrices ${\bf P}%
^{\prime A}$ and ${\bf P}^{\prime B}$,
\begin{eqnarray}
G_{jl}=\left\langle \Psi _F^A|\Psi _F^B\right\rangle =\det \left[
\left( {\bf P}^{\prime A}\right) ^T{\bf P}^{\prime B}\right] .
\end{eqnarray}
Using the anyonic commutation relation the ROBDM can be expressed as
\begin{eqnarray}
\rho _{jl}=\left\langle a_j^{\dagger }a_l\right\rangle =\delta
_{jl}\left( 1-G_{jl}\right) -(1-\delta_{jl})e^{-i\chi \pi }G_{jl}.
\end{eqnarray}
The natural orbitals $\phi ^{\eta}$ are defined as the
eigenfunctions of the one-particle density matrix,
\begin{eqnarray}
\sum_{j=1}^L\rho _{jl}\phi ^{\eta}=\lambda _{\eta}\phi ^{\eta},
j=1,2, ..., L,
\end{eqnarray}
and can be understood as the effective single-particle states with
occupations $\lambda _{\eta}$. In order to investigate the effect of
statistical parameter $\chi$ we will focus on the momentum
distribution, which is defined as
\begin{eqnarray}
n(k)=\frac 1{2\pi}\sum_{j,l=1}^Le^{-ik(j-l)}\rho _{jl}.
\end{eqnarray}

\section{properties of the ground state}

\begin{figure}[tbp]
\includegraphics[width=3.5in]{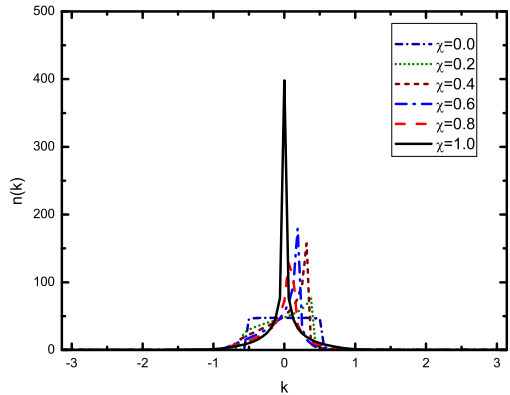}\newline
\caption{(Color online) Momentum distribution of the ground state
for $V_0=0$, $ N=50$, and $L=300.$ The unit of $k$ is $1/a$ with $a$
the lattice constant. For convenience, hereafter we set $a=1$.}
\label{fig1}
\end{figure}

We first investigate the momentum distributions for the situation in
the absence of a harmonic trap. In this case the single-particle
eigenstate for $t=1.0$ can be formulated as
$\phi_{\alpha}(l)=\sin\alpha l \pi/(L+1)/\sqrt {C_\alpha}$ with the
normalized constant $C_\alpha=\sum_{j=1}^L\sin ^2\alpha j\pi/(L+1)$
for the fixed boundary condition \cite{DMRGBook}. Following the
procedure in Sec. II the momentum distribution can be obtained
exactly and is shown in Fig. 1. The similar behavior as the finite
continuum model is manifested \cite{HaoPRA78}. In the Bose limit
($\chi =1$) most of bosons populate at the zero momentum state,
which corresponds to a Bose superfluid, while in the Fermi limit
($\chi =0$) the step-function distribution is shown, which is the
characteristic feature of free fermions. Both of them are symmetric
about the zero momentum. When the statistical parameter $\chi$
deviates from these two limits the momentum distribution is
asymmetric about the zero momentum. Anyons distribute in the regime
of positive momentum with more probability than that in the regime
of negative momentum and redistribute between these two regimes with
the increase in statistical parameter $\chi$. Finally the
distribution evolves between Fermi and Bose distribution with the
change in statistical parameter.

\begin{figure}[tbp]
\includegraphics[width=3.8in]{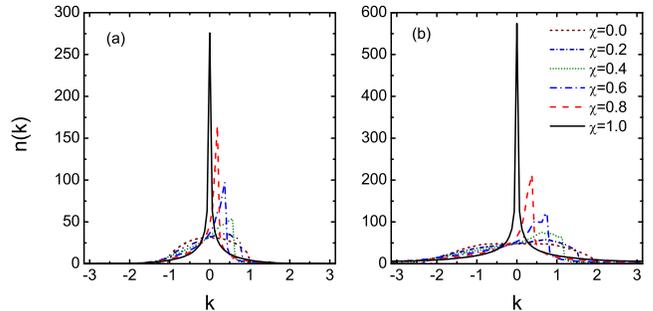}\newline
\caption{(Color online) Momentum distribution of the ground state
for $V_0=1.0\times 10^{-4}t$, $N=50$ (a) and (b) 150, and $L=300.$}
\label{fig2}
\end{figure}
When the harmonic trap is present we have to turn to the numerical
diagonal method \cite{MRigol} in order to get the $N$ lowest single
particle eigenstates of free spinless Fermi model and construct the
ground state of many body anyonic system. The momentum distributions
are displayed in Fig. 2 for filling number $N=50$ and $N=150$
confined in an optical lattice with $L=300$ sites. In this situation
the anyons populate in broader regime compared with the
confinement-free case as shown in Fig. 1. This is more evidently
seen in the Fermi limit. In the absence of harmonic trap anyons
(pure Fermions in this limit) occupy the momentum states in the
regime of $-1/2<k<1/2$ almost homogeneously while the combined
harmonic trap results that higher momentum states are occupied and
the distribution is no longer a sharp step-function. For all
statistical parameters the half width of momentum distribution
becomes wider. This can be understood by uncertainty principle that
the presence of harmonic trap reduces the uncertainty in the
coordinate space and leads to the increase of uncertainty in the
momentum space. When more particles are loaded in the lattice they
will populate at higher momentum states with more probability and
the peak of momentum profiles will not be so sharp as those in the
case of less filling number. Obviously the filling number and the
strength of harmonic trap shall not affect the statistical property
that with the change in statical parameter anyon's momentum
distribution evolves continuously from a Bose distribution to a
Fermi one. In these two limits the distributions are symmetry about
the zero momentum and in between the profiles are asymmetric.

\begin{figure}[tbp]
\includegraphics[width=4.0in]{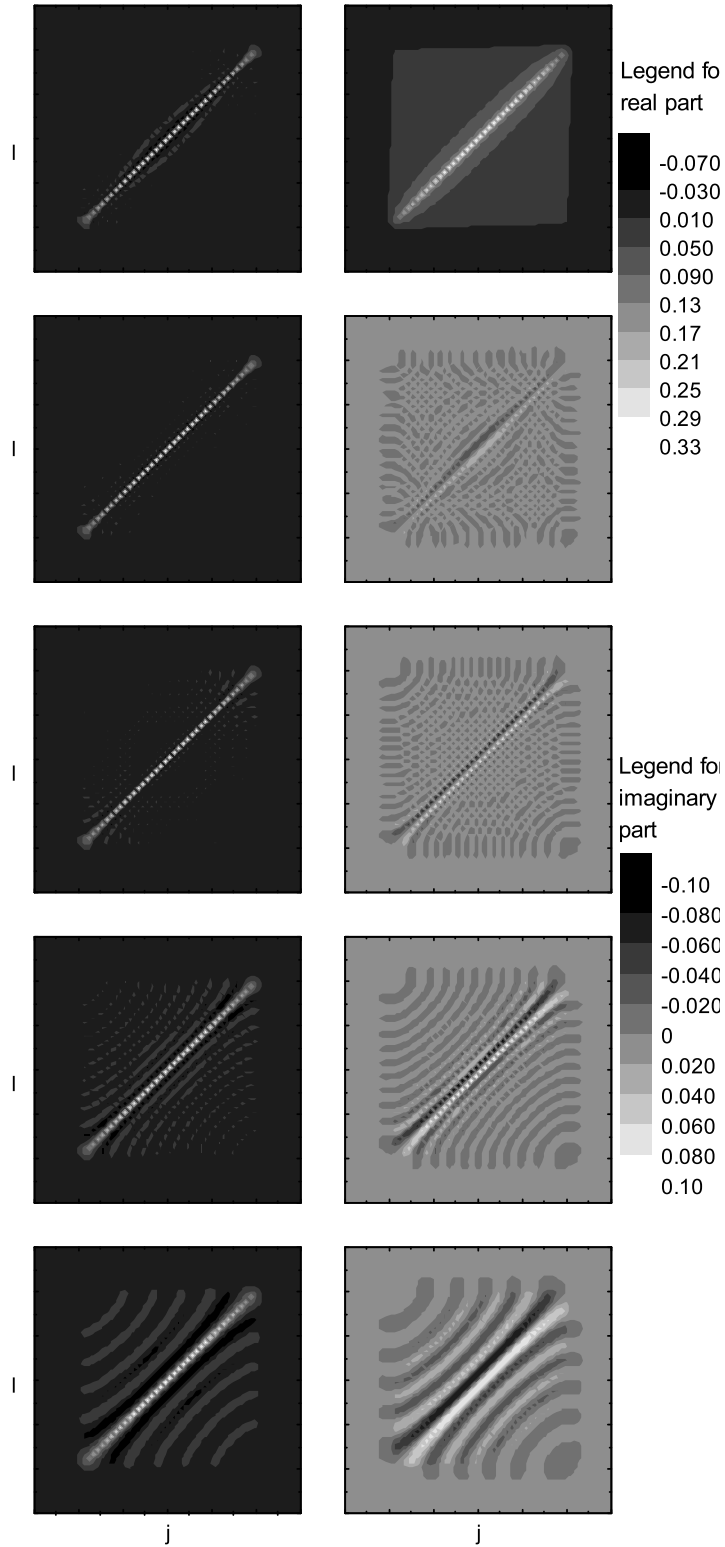}\newline
\caption{ Density matrix $\rho _{jl}$ for trapped systems with 300
lattice sites, $V_0=1.0\times 10^{-4}t$, and occupations of 50
HCA¡¯s. $j$ and $l$ range from 1 to 300. Top row: $\chi = 0.0$
(left) and 1.0 (right); Bottom four rows: $\chi = 0.2, 0.4, 0.6,
0.8$ (from top to bottom) for real part (left panel) and imaginary
part (right panel).} \label{fig3}
\end{figure}

In order to clarify the origin of asymmetric momentum distribution
of anyons we show the ROBDM for $L=300$ lattice sites with 50 anyons
in Fig. 3, where $j$ and $l$ ranged from 1 to $L$. In the Bose and
Fermi limits the ROBDMs are real (top row) whereas in the regime
deviating from these two limits the ROBDMs are Hermitian, i.e.,
possessing symmetric real part
(Re$\left[\rho_{jl}\right]$=Re$\left[\rho_{lj}\right]$) and
anti-symmetric imaginary part
(Im$\left[\rho_{jl}\right]$=-Im$\left[\rho_{lj}\right]$), which are
exhibited in last four rows for $\chi$ =0.2, 0.4, 0.6, and 0.8.
Therefore the formula of momentum distribution can be departed into
two parts as below that one part is even function of $k$ and the
other one is odd function of $k$,
\begin{eqnarray}
n\left( k\right)
&=&\frac1{2\pi}\sum_{j,l=1}^Le^{-ik(j-l) }\rho_{jl} \\
&=&\frac1{2\pi}\sum_{j,l=1}^L\left\{ \mathop{\rm Re} \left[ \rho
_{jl}\right] \cos k\left( j-l\right) + \mathop{\rm Im} \left[ \rho
_{jl}\right] \sin k\left( j-l\right) \right\} .\nonumber
\end{eqnarray}
It is because of this that the momentum distribution becomes
asymmetric about zero momentum. The imaginary part Im$\rho_{jl}$ is
an odd function of statistical parameter $\chi$ so the peak at
positive momentum as shown above will shift to negative momentum if
we take $\chi$ as negative ($\chi \rightarrow - \chi$)
\cite{HaoPRA78}.

In Fig. 4a we display the occupation of natural orbitals for the
system with 50 anyons in 300 lattice sites combined with weak
harmonic trap ($V_0=1.0\times 10^{-4}t$). In the Fermi limit each
anyon occupies one orbital and the occupation distribution is a step
function. In the Bose limit most anyons occupy the lower orbitals
and the occupation distribution exhibits sharp single-peak
structure. In the regime deviating from these two extreme points the
evolution from one to the other indicates that with the decrease in
statistical parameter more and more anyons occupy higher orbitals
and finally distribute in the lowest $N$ orbitals homogeneously in
the Fermi limit. The statistical effect on the natural orbital is
shown in Figs. 4b-4d for the same system as Fig. 4a, where the
modulus of the lowest natural orbitals ($|\phi^1|$) is exhibited for
statistical parameter $\chi$=0.0, 0.4, and 1.0. In the Bose limit
the modulus are almost smooth and only weak oscillations appear
around the boundary while the oscillations become more and more
obvious with the decrease in statistical parameter. In the Fermi
limit $N$ peaks display clearly.
\begin{figure}[tbp]
\includegraphics[width=3.3in]{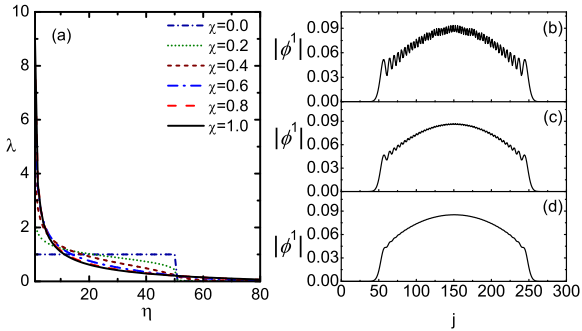}\newline
\caption{(Color online) (a) Occupation of the natural orbitals; the
modulus of the lowest natural orbital for (b) $\chi=0.0$, (c)
$\chi=0.4$, and (d) $\chi=1.0$. } \label{fig4}
\end{figure}

\section{conclusions}

In summary, we have investigated the ground-state properties of 1D
anyon gas confined in optical lattices combined with a weak harmonic
trap in the hard core limit using exact numerical method. With
Jordan-Wigner transformation the hard-core anyon model is related to
polarized free spinless Fermi model and thus exact ground state can
be constructed from that of $N$ free fermions which is composed of
the lowest $N$ single-particle eigenstates of free fermion. Then by
calculating the one-particle Green's function we obtain the ROBDM
and momentum distribution. It is indicated that the ROBDM is a
complex Hermitian matrix and the momentum distributions show
properties distinct from the bosons and fermions. In the Bose limit
and Fermi limit the ROBDM is real and the momentum distributions are
symmetric about the zero momentum while anyons populate in the
momentum space asymmetrically and would rather stay in some special
regime with large probability. With the change in statistic
parameter the system exhibits the Bose statistics, Fermi statistics
and the fractional statistics in between. The anyonic system
exhibits characteristic feature between Bose and Fermi statistics
also in the occupation of natural orbitals and the modulus of
natural orbital shows more and more obvious oscillation with the
decrease in statistical parameter.

\begin{acknowledgments}
This work was supported by NSF of China under Grants  No. 10821403,
No. 10774095, and No. 10847105, the 973 Program under Grant No.
2006CB921102, and National Program for Basic Research of MOST,
China.

\end{acknowledgments}

\end{document}